\documentclass[12pt]{amsart}
\usepackage{amssymb}

\newcommand{\wt}[1]{\widetilde{#1}}
\newcommand{\wh}[1]{\widehat{#1}}

\newtheorem{thm}{Theorem}

\begin{document}

\title[Root systems and Kochen-Specker theorem]{Exceptional and non-crystallographic 
root systems and the Kochen-Specker theorem}

\author[A. Ruuge]{Artur E. Ruuge}
\address{
Department of Quantum Statistics and Field Theory, 
Faculty of Physics, Moscow State University, 
Vorobyovy Gory, 119899, Moscow, Russia 
\emph{and} 
Faculteit Ingenieurswetenschappen, 
Vrije Universiteit Brussel, 
Pleinlaan 2, 
B-1050 Brussel, Belgi\"e
}
\email{Artur.Ruuge@ua.ac.be} 

\begin{abstract}
The Kochen-Specker theorem states that a 3-dimen\-sional complex Euclidean space admits a 
finite configuration of projective lines such that the corresponding quantum observables 
(the orthogonal projectors) cannot be assigned with 0 and 1 values in a classically consistent way. 
This paper shows that the irreducible root systems of exceptional and of non-crystallographic types 
are useful in constructing such configurations in other dimensions. 
The cases $E_6$ and $E_7$ lead to new examples, while $F_4$, $E_8$, and $H_4$, 
yield a new interpretation of the known ones.   
The described configurations have an additional property: they are saturated, i.e. 
the tuples of mutually orthogonal lines, being partially ordered by inclusion, yield a poset 
with all maximal elements having the same cardinality (the dimension of space).

\end{abstract}

\maketitle

\section{Introduction}

The aim of the present paper is to establish a link between 
several examples illustrating the Kochen-Specker theorem \cite{KochenSpecker} 
(a result in non-relativistic quantum theory closely related to Bell's inequalities)
and the irreducible root systems (a notion emerging in the classification of finite-dimensional 
complex simple Lie algebras and of finite Coxeter groups). 

Let us recall what the Kochen-Specker result is about. 
The main object is a \emph{finite} collection 
$A \subset \mathbb{P} (\mathcal{H})$ 
of (projective) lines  in a 
complex or real Hilbert space $\mathcal{H}$ of \emph{finite} dimension $d$. 
One is interested in the orthogonality relation $\perp$ between the elements of $A$, and the 
aim is to assign to each $x \in A$ one of two colors, say \emph{red} or \emph{blue}, 
satisfying certain conditions.   
Every such (bi)colouring is described by a function $v : A \to \lbrace 0, 1 \rbrace$, where 
$0$ corresponds, say, to the blue colour and $1$ corresponds to the red colour. 
Let us say that a bicolouring $v: A \to \lbrace 0, 1 \rbrace$ is 
\emph{good} if 
(1) for all collections of mutually orthogonal $x_1, x_2, \dots, x_d \in A$
there exists a unique $i_0$ such that $v (x_{i_0}) = 1$; 
(2) $\forall x, y \in A$: if 
$y \perp x$ and $v (x) = 1$, then $v (y) = 0$. 
S.Kochen and E.P.Specker prove that if $d = 3$ and $\mathcal{H}$ is complex, then 
there exists $A$ which does not admit a good $v$ 
(note that their construction is explicit and yields $| A | = 117$).  

Let us call $A$ \emph{non-colourable} if it does not admit 
good bicolourings, and \emph{colourable} -- otherwise (leaving out the prefix \emph{bi}). 
The motivation to look for such configurations  
comes from the analysis of quantum theory in terms of Bell's inequalities 
meant to express the deviation of the behavior of  
a physical system from a classical pattern. 
If each $x \in A$ is identified with a $0$-$1$ observable (represented 
by the orthogonal projector on $x$), then the behavior of the physical 
system with respect to the measurement acts of these observables will be 
extremely non-classical in the following sense. 
Take any $A$ with the property mentioned. 
In classical physics an act of measurement of an observable is an act of 
\emph{revealing} of its pre-existing value. 
Try to accept 
the same point of view in the quantum case, in particular, for the observables $A$.   
Suppose we have a collection of mutually orthogonal $x_1, x_2, \dots x_k \in A$. 
Since the corresponding projectors commute, their observables can be measured simultaneously. 
If a simultaneous measurement act yields 1 for $x_{i_0}$, 
it yields $0$ for the other $x_{i}$, $i \not = i_0$. 
If $k = d$, then $x_{i_0}$ corresponding to $1$ is always present.  
Therefore we can induce a \emph{good} bicolouring $v$ on $A$ 
with $v (x)$ being the pre-existing value corresponding to $x \in A$.  
Since $A$ \emph{does not} admit such bicolourings, one may not 
interpret the acts of measurements with respect to $A$ in a straightforward classical fashion. 

The non-colourable configurations $A$ are known to exist 
in every dimension $d \geqslant 3$. 
For $d \leqslant 2$ all projective configurations admit good (bi)colourings. 
It turns out that some of the examples have nice geometrical properties. 
If we look at the whole collection $\mathbb{P} (\mathcal{H})$, 
$\mathrm{dim} \mathcal{H} = d$, 
then we have the following: 
whenever $x_1, x_2, \dots, x_k \in \mathbb{P} (\mathcal{H})$, $k < d$, are mutually orthogonal, 
there exist $x_{k+1}, x_{k + 2}, \dots, x_{d}$, such that 
$x_1, x_2, \dots, x_d$ are mutually orthogonal. 
Let us require this property from a configuration $A \subset \mathbb{P} (\mathcal{H})$. 
Denote 
\begin{equation*} 
\mathcal{P}_{\perp}^{(k)} (A) := \lbrace 
U \subset A \, | \, \# U = k \, \& \, \forall x, y \in U: x \not = y \Rightarrow x \perp y 
\rbrace. 
\end{equation*} 
Put $\mathcal{P}_{\perp} (A) := \cup_{k = 0}^{d} \mathcal{P}_{\perp}^{(k)} (A)$. 
$A$ is called \emph{saturated} if $\forall U \in \mathcal{P}_{\perp} (A)$ 
$\exists M \in \mathcal{P}_{\perp}^{(d)} (A)$ such that $M \supset U$. 
An easy example of a finite saturated configuration $A$ in $d$ dimensions is just a collection of 
$d$ mutually orthogonal lines, but there exist much more complicated examples. 
Furthermore, there exists finite saturated configurations which 
do not admit good bicolourings!
From a quantum mechanical perspective, one may 
view such configurations as finite analogs of 
$\mathbb{P} (\mathcal{H})$.

Intuitively, a finite saturated projective configuration 
without good bicolourings is something very symmetric. 
This symmetry is essentially the subject of the present paper.  
It turns out that exceptional root systems and non-crystallographic root systems of finite 
Coxeter groups allow to construct examples of such configurations. 
The idea is to consider the projective lines represented by the roots. 
There exist the following exceptional root systems: 
$G_2$, $F_4$, $E_6$, $E_7$, and $E_8$. 
The non-crystallographic root systems are denoted by $I_2 (p)$ ($p = 5$ or $p > 6$), $H_3$, and $H_4$. 
Since a configuration is non-colourable only if $d \geqslant 3$, 
focus on the root systems $F_4$, $E_6$, $E_7$, $E_8$, $H_3$, and $H_4$.  
The result is as follows: 

\begin{thm} 

\begin{itemize} 
\item[(1)] 
The finite projective configurations $F_4$, $E_7$, $E_8$, and $H_4$ 
are saturated and non-colourable. 

\item[(2)]
The configuration $H_3$ is saturated, but colourable. 

\item[(3)] 
The configuration $E_6$ admits an extension up to a saturated finite configuration, 
which is non-colourable.

\end{itemize}
\end{thm}

The explicit description of the mentioned saturation of $E_6$ 
configuration will be given below (theorem 2). 
It is interesting to mention, that it is realized by a presheaf-like construction that 
makes use of the remaining exceptional root system $G_2$.

\section{The root systems $F_4$, $E_8$, $H_4$ and $H_3$.} 

The $F_4$, $E_8$ and $H_4$ configurations correspond to the 
Kochen-Specker-type examples already considered in the literature. 
Therefore we make just a few remarks. 
The saturation property can be verified on a personal computer in 
a straightforward manner (for example, in Maple). 

The projective configurations in $\mathbb{R}^4$ illustrating the 
Kochen-Specker theorem 
given by A.Peres \cite{Peres} (20 lines) and 
A.Cabello, J.M.Esterbaranz, G.Garc\'ia-Alcaine \cite{Cabello} (18 lines) 
can be viewed as subsets of the same set of 24 lines represented by 
the elements of the $F_4$ root system. 

The root system $E_8$ is related to the Kochen-Specker-type example constructed by D.Mermin \cite{Mermin} and by 
M.Kernaghan, A.Peres \cite{KernaghanPeres}. Their example involves 40 projective lines in $\mathbb{R}^8$. 
The finite \emph{saturated} configuration containing these lines has been 
constructed by A.Ruuge, F.Van Oystaeyen \cite{RuugeFVO1}. 
It consists of 120 projective lines. 
These lines can be viewed as projective lines 
corresponding to the 240 roots of the irreducible root system $E_8$. 

The $H_4$ case corresponds to the paper of P.K.Aravind, F.Lee-Elkin \cite{Aravind}. 
There are 120 roots, which yield 60 projective lines in $\mathbb{R}^4$. 

The $H_3$ case is rather simple and does not yield a new example of a finite non-colourable 
configuration (in $\mathbb{R}^3$). 
The root system contains the vectors 
$(\pm 1, 0, 0)$, $(\pm 1, \pm 1/ \tau, \pm \tau)$, 
plus all the vectors obtained from them by cyclic permutations of coordinates; 
here $\tau = (1 + \sqrt{5})/ 2$ is the \emph{golden ratio} (recall that 
$\tau^2 = \tau + 1$). There are 30 roots in total, and therefore 15 projective lines. 
The corresponding configuration uniquely splits into five mutually disjoint triples 
of mutually orthogonal lines. It is saturated and colourable, admitting $3^5$ good bicolourings.

\section{The root system $E_7$.}

The case of the root system of the Coxeter group of type $E_7$ 
requires a little bit more work. 
It is necessary to make some remarks to persuade oneself in 
the fact that the corresponding (saturated) configuration (in $\mathbb{R}^7$) is not 
colourable. 

It is convenient to model the $E_7$ root system (denote it by $\Phi$)
not on $\mathbb{R}^7$, but on a 7-dimensional subspace of $\mathbb{R}^8$ 
consisting of all vectors $(a_1, a_2, \dots, a_8)$ such that $\sum_{i = 1}^{8} a_i = 0$. 
One can obtain $\Phi$ by taking the 
union of the orbits of 
$(1, \bar 1, 0, 0, 0, 0, 0, 0)$ and 
$(1/2) \, (1, 1, 1, 1, \bar 1, \bar 1, \bar 1, \bar 1)$
under the natural action of $S_8$ on $\mathbb{R}^8$; here $\bar 1 = - 1$. 
The result is $|\Phi| = 126$ and therefore we have $63$ projective lines (rays). 
Let $[a_1, a_2, \dots, a_8]$ denote the ray represented by the vector $(a_1, a_2, \dots, a_8)$. 

For each $k = 2, 3, \dots, 7$, one may consider all $k$-tuples of mutually orthogonal rays; 
denote the number of all such $k$-tuples by $n_k$. 
Then a (straightforward) Maple computation yields: 
$n_2 = 945$, 
$n_3 = 4095$, 
$n_4 = 4725$,  
$n_5 = 2835$,  
$n_6 = 945$,  
$n_7 = 135$. 
It turns out, that our configuration (63 rays) can be represented as a 
union of nine mutually disjoint 7-tuples of mutually orthogonal rays ($63 = 9 \times 7$). 
In fact, there are 960 possibilities to realize such a disjoint union, but 
we are going to select just one of them. To describe it, it is convenient to index the components of 
$a \in \mathbb{R}^8$ not by $1, 2, \dots, 8$, but by the elements of the projective line 
$\mathbb{F}_{7} \cup \lbrace \infty \rbrace$ over the field of 7 elements, 
$a = (a_{\infty}, a_0, a_1, \dots, a_6)$. 

Let $k$ vary over $\mathbb{F}_7$. Denote by $\lambda^{(k)}$ the ray represented by 
the vector $(a_{\infty}, a_0, \dots, a_6)$ 
having $a_{\infty} = 1$, $a_k = -1$, and all other components equal to $0$. 
Denote by $\mu^{(k)}$ the ray represented by the vector having 
$1$ at the positions $\infty, k, k + 1, k + 3$, and $- 1$ at the other four positions. 
Denote by $\nu^{(k)}$ the ray represented by the vector having 
$1$ at the positions $\infty, k, k - 1, k - 3$, and $- 1$ at the other four positions. 
Next, let $i \in \mathbb{F}_7$ vary over $1, 2, 3$. 
Denote by $\xi^{(k, i)}$ the ray represented by the vector
$(a_{\infty}, a_0, \dots, a_6)$ 
having $a_{k + i} = 1$, $a_{k - i} = -1$, and all other components equal to $0$. 
Finally, denote by $\eta^{(k, i)}$ the ray represented by the vector 
having $1$ at the positions $\infty, k, k + i, k - i$, and $-1$ at the other four positions. 

With this notation we can describe the following 7-tuples of mutually orthogonal rays. 
Put $Q_k := \lbrace \lambda^{(k)}, 
\xi^{(k, 1)}, \xi^{(k, 2)}, \xi^{(k, 3)}, 
\eta^{(k, 1)}, \eta^{(k, 2)}, \eta^{(k, 3)}
\rbrace$, $k \in \mathbb{F}_7$. 
Put $Q_{+} := \lbrace \mu^{(0)}, \mu^{(1)}, \dots, \mu^{(6)}\rbrace$ and 
$Q_{-} := \lbrace \nu^{(0)}, \nu^{(1)}, \dots, \nu^{(6)}\rbrace$. 
It is straightforward to check that $Q_0, \dots, Q_6, Q_{+}, Q_{-}$ are 
mutually disjoint; their union is just the $E_7$ configuration. 
Let us also write $Q_7$ instead of $Q_{+}$, and $Q_{8}$ instead of $Q_{-}$. 

The verification that the configuration is non-colourable can now be completed on a computer. 
If it were colourable, one could choose in each $Q_i$ an element $l_i$, in such a way, that 
the rays $l_1, l_2, \dots, l_9$ were pairwise \emph{non-orthogonal}. 
To verify that this is impossible, take any $x_1 \in Q_1$. 
Find $x_2 \in Q_2$ such that $x_2 \not \perp x_1$. After that, 
find $x_3 \in Q_3$ such that $x_3 \not \perp x_1$ and $x_3 \not \perp x_2$. 
If this is possible, try to find $x_4 \in Q_4$ such that $x_4 \not \perp x_1, x_2, x_3$, and so on. 
It turns out, that one cannot reach this way the set $Q_9$. 
Therefore the $E_7$ configuration is non-colourable. 
Introducing (in analogy with $n_k$) the numbers $m_q$ for the numbers of all 
$q$-tuples of mutually \emph{non-orthogonal} rays, one obtains: 
$m_2 = 1008$, $m_3 = 5376$, $m_4 = 10080$, $m_5 = 8064$, $m_6 = 2016$, $m_7 = 288$, $m_8 = 0$. 
An example of seven mutually non-orthogonal rays is 
$\lambda^{(0)}, \lambda^{(1)}, \dots, \lambda^{(6)}$.   

\section{The root system $E_6$.}

This case is much more complicated than the other cases. 
The corresponding configuration does not contain tuples of pairwise orthogonal rays 
which have 5 or 6 elements. 
In particular, it is not saturated. 
It turns out that it is possible to construct a non-colourable saturated configuration containing it.  
Moreover, the construction makes use of the remaining $G_2$ root system, i.e. 
in the end \emph{all} exceptional root systems turn out to be useful in constructing 
the examples of non-colourable saturated configurations. Generally speaking, what happens is that 
one computes all the 4-tuples of pairwise orthogonal lines in $E_6$, and then attaches 
to each such tuple a copy of $G_2$ projective configuration. This presheaf-like construction 
turns out to be saturated. Some more lines are needed to achieve non-colourability, but the saturation 
property can be preserved.

Let us describe the roots of $E_6$. 
It is convenient to model them on a 6-dimensional subspace $R$ of $\mathbb{R}^9$. 
A generic element of $\mathbb{R}^9$ is of the form 
$(x_1, x_2, x_3; y_1, y_2, y_3; z_1, z_2, z_3)$. 
It is convenient to use a shorter notation for it: $(x; y; z)$, where 
$x = (x_1, x_2, x_3)$, $y = (y_1, y_2, y_3)$, and $z = (z_1, z_2, z_3)$. 
The conditions defining $R$ are: 
$x_1 + x_2 + x_3 = 0$, $y_1 + y_2 + y_3 = 0$, and $z_1 + z_2 + z_3 = 0$. 

The root system contains 72 elements. 
Some of the vectors are of the form $(\xi; \theta; \theta)$, 
$(\theta; \xi; \theta)$, and $(\theta; \theta; \xi)$, 
where $\theta = (0, 0, 0)$, and $\xi = (1, \bar 1, 0)$, $(1, 0, \bar 1)$, or $(0, 1, \bar 1)$, 
$\bar 1 \equiv -1$.  
This yields 9 elements. 
The other part of elements is given by the triples 
$(\xi; \eta; \zeta)$, where $\xi$, $\eta$, $\zeta$ vary over $\lbrace (1/ 3) (2, \bar 1, \bar 1), 
(1/ 3) (\bar 1, 2, \bar 1), (1/ 3) (\bar 1, \bar 1, 2) \rbrace$. 
This yields 27 elements; in total we obtain 36 elements. 
The remaining 36 elements of the root system are just the inverses of the described ones.  

The 72 roots define 36 projective lines (rays). Denote this set by $A$. 
The configuration is quite symmetric: each line is orthogonal to precisely 15 other lines. 
Note, that each line can be represented by an integer (9-dimensional) vector with the 
entries being $0$, $\pm 1$, or $\pm 2$. 
Denote by $n_m$ the number of subsets $U \subset A$ consisting of $m$ pairwise orthogonal lines. 
A computation in analogy with the $E_7$ case yields  
$n_2 = 270$, 
$n_3 = 540$, 
$n_4 = 135$, 
$n_5 = 0$, 
$n_6 = 0$. 
There are no tuples of cardinality 5 and 6, but for smaller tuples  
one can check, that each pair of orthogonal lines extends to a triple, and each triple 
extends to a 4-tuple (of pairwise orthogonal lines). 
In this sense the configuration ``tries to be saturated''.

We need more lines to construct a 6-dimensional saturated configuration. 
Look at the 4-tuples of pairwise orthogonal lines in $A$. 
They can be classified.  
There are tuples of the form:
\begin{equation*} 
\begin{aligned}
Q_1 := \lbrace 
&[1, \bar 1, 0; \quad 0, 0, 0; \quad 0, 0, 0], \\
&[0, 0, 0; \quad 1, \bar 1, 0; \quad 0, 0, 0], \\
&[0, 0, 0; \quad 0, 0, 0; \quad 1, \bar 1, 0], \\
&[1, 1, \bar 2; \quad 1, 1, \bar 2; \quad 1, 1, \bar 2] \rbrace,  
\end{aligned}
\end{equation*}
where the bar denotes negation. 
Similar tuples are obtained by permutations of coordinates. 
There are 27 tuples of this type. 

The other type of tuples is represented by 
\begin{equation*}
\begin{aligned}
Q_2 := \lbrace 
&[1, \bar 1, 0; \quad 0, 0, 0; \quad 0, 0, 0], \\
&[1, 1, \bar 2; \quad \bar 2, 1, 1; \quad \bar 2, 1, 1], \\
&[1, 1, \bar 2; \quad 1, \bar 2, 1; \quad 1, \bar 2, 1], \\
&[1, 1, \bar 2; \quad 1, 1, \bar 2; \quad 1, 1, \bar 2]
\rbrace. 
\end{aligned}
\end{equation*}
Permutations of coordinates yield 54 different tuples of this form. 

The third type of tuples is represented by 
\begin{equation*}
\begin{aligned}
Q_3 := \lbrace
&[\bar 2, 1, 1; \quad \bar 2, 1, 1; \quad \bar 2, 1, 1], \\
&[\bar 2, 1, 1; \quad 1, \bar 2, 1; \quad 1, \bar 2, 1], \\
&[1, \bar 2, 1; \quad \bar 2, 1, 1; \quad 1, \bar 2, 1], \\
&[1, \bar 2, 1; \quad 1, \bar 2, 1; \quad \bar 2, 1, 1]
\rbrace. 
\end{aligned}
\end{equation*}
Permutations yield again 54 different variants. 
In total we have $54 + 54 + 27 = 135$ ($n_4 = 135$) tuples. 
 
Now look at the subspaces (of the 6-dimensional space $R$ mentioned) \emph{orthogonal} to these 4-tuples. 
Let $\xi, \eta, \zeta$ be real variables satisfying  $\xi + \eta + \zeta = 0$.
A generic element of $Q_{1}^{\perp} \cap R$ can be written as 
$[\xi (1, 1, \bar 2); \, \eta (1, 1, \bar 2); \, \zeta (1, 1, \bar 2)]$. 
A generic element of $Q_{2}^{\perp} \cap R$ is of the form 
$[0, 0, 0; \, \xi, \eta, \zeta; \, - \xi, - \eta, - \zeta]$. 
The space $Q_{3}^{\perp} \cap R$ coincides with $Q_{1}^{\perp} \cap R$.

Invoke the exceptional root system $G_2$. 
Its roots are naturally modeled on a 2-dimensional subspace 
$x + y + z = 0$ of the space of 3-dimensional vectors $(x, y, z)$. 
The roots are: $(1, \bar 1, 0)$, $(1, 0, \bar 1)$, $(0, 1, \bar 1)$, 
$(2, \bar 1, \bar 1)$, $(\bar 1, 2, \bar 1)$, $(\bar 1, \bar 1, 2)$, and their inverses (i.e. there are 12 roots). 
The idea is to identify this subspace with the 2-dimensional subspaces defined by the parameters 
$\xi, \eta, \zeta$. 
In other words, enhance the $E_6$ projective configuration with the projective lines 
represented by such vectors, for which $(\xi, \eta, \zeta)$ is an element of $G_2$ root system. 
For example, if we take the 4-tuple $Q_1$, we obtain 6 projective lines of the form 
$[\xi (1, 1, \bar 2); \, \eta (1, 1, \bar 2); \, \zeta( 1, 1, \bar 2)]$, where 
$(\xi, \eta, \zeta)$ varies over $(1, \bar 1, 0)$, $(1, 0, \bar 1)$, $(0, 1, \bar 1)$, 
$(2, \bar 1, \bar 1)$, $(\bar 1, 2, \bar 1)$, $(\bar 1, \bar 1, 2)$. 

Recall that we have 135 different 4-tuples of pairwise orthogonal 
projective lines corresponding to the $E_6$ root system. 
Construct for each such tuple the six projective lines (invoking the $G_2$ root system). 
Take the union of all these lines. This yields 162 lines. 
Attaching them to $A$, we obtain a projective configuration $\wt{A} \supset A$ of 198 elements. 

It turns out that $\wt{A}$ is saturated! 
We can generate the $m$-tuples of pairwise orthogonal lines of $\wt{A}$ .  
Let $\wt{n}_m$ be the number of these tuples. 
The Maple computation results in  
$\wt{n}_2 = 4995$, 
$\wt{n}_3 = 25920$, 
$\wt{n}_4 = 32400$, 
$\wt{n}_5 = 15552$, 
$\wt{n}_6 = 2592$. 
One could hope that $\wt{A}$ is non-colourable, but the situation is slightly more complicated. 
It turns out (see below) that $\wt{A}$ is colourable, but there is \emph{only one} good bicolouring. 
This immediately leads to the idea of how to construct a non-colourable configuration 
containing $E_6$. Since two red rays cannot be orthogonal (by the definition of a good 
bicolouring), one can consider a copy $\wt{A}'$ of $\wt{A}$ obtained by some rotation. 
It is possible to adjust this rotation in such a way that at least one of the red rays in $\wt{A}$ is 
orthogonal to a red ray in $\wt{A}'$. Then $\wt{A} \cup \wt{A}'$ becomes non-colourable. 
Furthermore (see below), one can choose this rotation in such a way that there exists a saturated finite 
configuration $\wh{A} \supset \wt{A} \cup \wt{A}'$. By that one arrives at a 
finite saturated non-colourable configuration $\wh{A}$ containing the $E_6$ configuration. 

Let us formulate the final result first and then give some comments. 
We have a 9-dimensional space $\mathbb{R}^9$ consisting of vectors 
$(x; y; z)$, where $x = (x_1, x_2, x_3)$, $y = (y_1, y_2, y_3)$, $z = (z_1, z_2, z_3)$. 
The symmetric group $S_3$ naturally acts in four different ways on these vectors: 
permuting $\lbrace x_i \rbrace_i$, $\lbrace y_j \rbrace_j$, $\lbrace z_k \rbrace_k$, or 
permuting $\lbrace x, y, z \rbrace$. 
This gives an action of the wreath product $S_3 \wr S_3$ on $\mathbb{R}^9$ which fixes $R$ 
(recall that $R$ is defined by the conditions $\sum_i x_i = \sum y_j = \sum_k z_k = 0$) 
on which we realize the $E_6$ root system. 
There are also three natural ways to act on $\mathbb{R}^9$ with the group $\mathbb{Z}/ 2 \mathbb{Z}$ 
by negating $x$, $y$, or $z$ components, respectively. This yields an 
action of $(\mathbb{Z}/ 2 \mathbb{Z})^3$ on $\mathbb{R}^9$, again fixing $R$. 
The constructed actions on $\mathbb{R}^9$ induce the actions on $\mathbb{P} (R)$ (the set of rays in $R$). 
For $\lambda \in \mathbb{P} (R)$, denote by $O (\lambda)$ its orbit under the action of $S_3 \wr S_3$, and 
by $\wh{O} (\lambda)$ the corresponding  orbit under the action of the 
free product of $S_3 \wr S_3$ and $(\mathbb{Z}/ 2 \mathbb{Z})^3$. 
Consider the following six rays: 
\begin{gather*} 
\lambda_{1} := [1, \bar 1, 0; \quad 0, 0, 0; \quad 0, 0, 0 ], \\
\lambda_{2} := [2, \bar 1, \bar 1; \quad 2, \bar 1, \bar 1; \quad 2, \bar 1, \bar 1], \\
\lambda_{3} := [1, \bar 1, 0; \quad 1, \bar 1, 0; \quad 0, 0, 0], \\
\lambda_{4} := [2, \bar 1, \bar 1; \quad \bar 2, 1, 1; \quad 0, 0, 0], \\
\lambda_{5} := [\bar 4, 2, 2; \quad 2, \bar 1, \bar 1; \quad 2, \bar 1, \bar 1], \\
\lambda_{6} := [2, \bar 1, \bar 1; \quad 0, 0, 0; \quad 0, 0, 0]. 
\end{gather*}
In this notation we have: 
\begin{thm} 

\begin{itemize} 
\item[(1)] 
The union 
$A := O (\lambda_1) \cup O (\lambda_2)$ yields the $E_6$
projective configuration;  

\item[(2)] 
The union $\wt{A} := \cup_{i = 1}^{5} O (\lambda_i)$ is a saturated projective configuration, 
$\wt{A} \supset A$, $| \wt{A} | = 198$. 
It admits precisely one good bicolouring. 
The set of red lines coincides with the orbit $O (\lambda_4)$; 

\item[(3)] 
The union $\wh{A} := \cup_{i = 1}^{6} \wh{O} (\lambda_i)$ is a saturated non-colourable 
projective configuration, $\wh{A} \supset \wt{A} \supset A$, $| \wh{A} | = 558$. 

\end{itemize}
\end{thm}

Let us describe the strategy of the implementation of the proof of this theorem on a computer. 
The facts that $\wt{A}$ and $\wh{A}$ are saturated can be verified in a straightforward way in Maple. 
The non-trivial part of the proof is to check that $\wt{A}$ admits just one good bicolouring.

The set $\wt{A}$ consists of two disjoint parts: $\wt{A} = A \cup A_{\mathrm{ext}}$, where 
$A$ is the set corresponding to $E_6$ roots. 
The first step will be to describe \emph{some} (not all) of the elements of 
$\mathcal{P}_{\perp}^{(6)} (\wt{A})$. These elements will be of the shape 
$T \cup B$, where $T \in \mathcal{P}_{\perp}^{(4)} (A)$ and $B \in \mathcal{P}_{\perp}^{(2)} (A_{\mathrm{ext}})$. 
More precisely, we shall describe a collection of $T_{p}^{(i)} \in \mathcal{P}_{\perp}^{(4)} (A)$ and 
$B_{p}^{(i)} \in \mathcal{P}_{\perp}^{(2)} (A_{\mathrm{ext}})$, where 
$p = 1, 2, \dots, 45$ and $i = 1, 2, 3$, such that for each $p$ one may combine 
any $T_{p}^{(l)}$ with any $B_{p}^{(j)}$ ($l, j = 1, 2, 3$) to obtain 
$T_{p}^{(l)} \cup B_{p}^{(j)} \in \mathcal{P}_{\perp}^{(6)} (\wt{A})$. 

Observe that there exist 6-tuples (of pairwise orthogonal lines) consisting just of the lines from 
$A_{\mathrm{ext}}$. 
For example: 
\begin{equation*} 
\begin{aligned}
P := \lbrace 
&[0, 0, 0; \quad 2, \bar 1, \bar 1; \quad \bar 2, 1, 1], \\
&[0, 0, 0; \quad 0, 1, \bar 1; \quad 0, 1, \bar 1], \\
&[0, 0, 0; \quad 0, 1, \bar 1; \quad 0, \bar 1, 1], \\
&[\bar 4, 2, 2; \quad 2, \bar 1, \bar 1; \quad 2, \bar 1, \bar 1], \\
&[2, \bar 4, 2; \quad 2, \bar 1, \bar 1; \quad 2, \bar 1, \bar 1], \\
&[2, 2, \bar 4; \quad 2, \bar 1, \bar 1; \quad 2, \bar 1, \bar 1]
\rbrace.
\end{aligned}
\end{equation*} 

Recall that $| \wt{A} | = 198$, $| A | = 36$. Therefore $| A_{\mathrm{ext}} | = 162$. 
Note, that 162 is divisible by 6; this is not an accident. 
Permuting the components of the 6 vectors (simultaneously), one can generate 
the 6-tuples similar to $P$. Each such tuple contains a unique element similar to 
$[0, 0, 0; \, 2, \bar 1, \bar 1; \, \bar 2, 1, 1]$ (i.e. the element in $O (\lambda_4)$). 
There are 27 tuples obtained this way. 
It is straightforward to check that they are mutually disjoint. 
Since $27 \times 6 = 162$, their union is just the set $A_{\mathrm{ext}}$.  

Let us number the mentioned 6-tuples of $A_{\mathrm{ext}}$ as 
$P_1, P_2, \dots, P_{27}$. 
Each of them contains exactly one point of $O (\lambda_4)$
(represented by the vector with precisely three zero coordinates); 
denote it by $a_i$, $i = 1, 2, \dots, 27$. 
Now look at the $A$ part of $\wt{A} = A \cup A_{\mathrm{ext}}$. 
Recall, that we have $n_4 = 135$ 4-tuples of pairwise orthogonal lines in $A$; 
number them in some way and denote as $T_{1}, T_{2}, \dots, T_{135}$. 
For each $i = 1, 2, \dots 27$, and each $b \in P_i \backslash \lbrace a_i \rbrace$, 
find all $m$, $1 \leqslant m \leqslant 135$, such that the elements of $T_m$ are orthogonal to 
$a_i$ and $b$. It turns out, that every time there are precisely 3 such 4-tuples. 
If $m_1 < m_2 < m_3$ are the three numbers of the tuples corresponding to $i$ and $b$, 
denote $\Delta_{i, b} := (m_1, m_2, m_3)$. 

Now consider the set $D := \lbrace \Delta_{i, b} \, | \, 
i = 1, 2, \dots 27; b \in P_i \backslash \lbrace a_i \rbrace\rbrace$. 
It's cardinality will be 45. For each $\Delta \in D$ compute all pairs of the form 
$(i, b)$, $1 \leqslant i \leqslant 27$, $b \in P_i \backslash \lbrace a_i \rbrace$, 
such that $\Delta_{i, b} = \Delta$.  
It turns out, that every time there are precisely three such pairs 
$(i, b)$, $(i', b')$, $(i'', b'')$ (let $i < i' < i''$); denote 
$G_{\Delta} := ((i, b), (i', b'), (i'', b''))$.  

Number the elements of $D$: $\Delta^{(1)}, \Delta^{(2)}, \dots, \Delta^{(45)}$. 
For each $p = 1, 2, \dots 45$, let $(m_{1}^{(p)},  m_{2}^{(p)}, m_{3}^{(p)}) = \Delta^{(p)}$; 
hence for each $p = 1, 2, \dots, 45$, we have 
three 4-tuples $T_{m_{1}^{(p)}}$, $T_{m_{2}^{(p)}}$, $T_{m_{3}^{(p)}}$ 
consisting of elements of $A$. 
From the corresponding $G_{\Delta^{(p)}}$, one obtains the three pairs 
$(a_{i}, b)$, $(a_{i'}, b')$, $(a_{i''}, b'')$ of elements of $A_{\mathrm{ext}}$. 
Redenote them $(u_p, v_p)$, $({u_p}', {v_p}')$, $({u_p}'', {v_p}'')$, 
respectively. 
For each $p = 1, 2, \dots, 45$, we have the data: 
\begin{equation*}
\sigma_{p} := (m_{1}^{(p)},  m_{2}^{(p)}, m_{3}^{(p)}; (u_p, v_p), (u_p', v_p'), (u_p'', v_p'')).  
\end{equation*}
Hence, the $27 \times 5 = 135$ distinct pairs of the form $(i, b)$, 
$i = 1, \dots 27$, $b \in P_i \backslash \lbrace a_i \rbrace$, 
are split into 45 triples $\lbrace (u_p, v_p), (u_p', v_p'), (u_p'', v_p'') \rbrace$, 
as well as the collection of 135 distinct 4-tuples $T_{m}$  
is split into 45 triples $\lbrace T_{m_{1}^{(p)}}, T_{m_{2}^{(p)}}, T_{m_{3}^{(p)}} \rbrace$.  
One may say that each $T$-triple is attached to a $(u, v)$-triple in $\sigma_{p}$: 
\begin{equation*} 
\lbrace T_{m_{1}^{(p)}}, T_{m_{2}^{(p)}}, T_{m_{3}^{(p)}} \rbrace 
\leftrightsquigarrow
\lbrace (u_p, v_p), (u_p', v_p'), (u_p'', v_p'') \rbrace. 
\end{equation*}
For every $p = 1, 2, \dots, 45$, 
combining a 4-tuple $T$ with any pair $(u, v)$, one obtains a tuple of six pairwise orthogonal 
lines in $A$. 
For example, $T_{m_{3}^{(p)}} \cup \lbrace u_p', v_p' \rbrace$ is a collection of 
six mutually orthogonal lines. 
As already mentioned, there exist other 6-tuples of pairwise orthogonal elements in $\wt{A}$, but 
it will suffice to consider just the described ones in order to establish the fact that  
there exists precisely \emph{one} good bicolouring of the set $\wt{A}$.

Now let us make the next step. 
For each $p = 1, 2, \dots, 45$, using the notation from the definition of $\sigma_{p}$, 
look at $s_p := \lbrace u_p, u_p', u_p'' \rbrace$. 
Recall, that $u_p$, $u_p'$, $u_p''$ are the projective lines of the form 
$a_i$, $1 \leqslant i \leqslant 27$. 
It turns out, that it is possible to \emph{partition} the set 
$\lbrace a_{1}, a_{2}, \dots, a_{27} \rbrace$ using the sets $\lbrace s_p \rbrace_p$ 
(note, that $27 = 3 \times 9$), i.e. there exists $(p_1, p_2, \dots, p_9)$ such that 
$s_{p_1}, s_{p_2}, \dots s_{p_9}$ are pairwise disjoint. 
As a remark, the Maple computation shows that 
each $s_p$ is disjoint with precisely 12 other sets of this form. 
Fix a concrete partition $P = (p_1, p_2, \dots, p_9)$ defined by 
\begin{gather*} 
s_{p_1} = \lbrace
[\theta; \xi_1; -\xi_2], 
[\theta; \xi_2; -\xi_3], 
[\theta; \xi_3; -\xi_1] 
\rbrace, \\
s_{p_2} = \lbrace
[\theta; \xi_1; -\xi_1], 
[\theta; \xi_2; -\xi_2], 
[\theta; \xi_3; -\xi_3]
\rbrace, \\
s_{p_3} = \lbrace
[-\xi_1; \theta; \xi_1], 
[-\xi_2; \theta; \xi_2], 
[-\xi_3; \theta; \xi_3]
\rbrace, \\
s_{p_4} = \lbrace
[-\xi_2; \theta; \xi_1], 
[-\xi_3; \theta; \xi_2], 
[-\xi_1; \theta; \xi_3]
\rbrace, \\
s_{p_5} = \lbrace
[\xi_1; -\xi_2; \theta], 
[\xi_2; -\xi_3; \theta], 
[\xi_3; -\xi_1; \theta]
\rbrace, \\
s_{p_6} = \lbrace
[\xi_1; -\xi_1; \theta], 
[\xi_2; -\xi_2; \theta], 
[\xi_3; -\xi_3; \theta]
\rbrace, \\
s_{p_7} = \lbrace 
[\theta; \xi_1; -\xi_3], 
[-\xi_2; \theta; \xi_1], 
[\xi_2; -\xi_1; \theta]
\rbrace, \\
s_{p_8} = \lbrace 
[\theta; \xi_3; -\xi_2], 
[-\xi_1; \theta; \xi_2], 
[\xi_1; -\xi_3; \theta]
\rbrace, \\
s_{p_9} = \lbrace
[\theta; \xi_2; -\xi_1], 
[-\xi_3; \theta; \xi_1], 
[\xi_3; -\xi_2; \theta]
\rbrace.
\end{gather*}

We have a collection of triples $(u_{p_i}, u_{p_i}', u_{p_i}'' )$, $i = 1, 2, \dots, 9$.  
For each triple we have $(v_{p_i}, v_{p_i}', v_{p_i}'')$ (see the notation in the definition of $\sigma_p$); 
$v_{p_i} \perp u_{p_i}$; $v_{p_i}' \perp u_{p_i}'$; $v_{p_i}'' \perp u_{p_i}''$.  
One may also consider 
$\Delta^{(p_i)} = (m_{1}^{(p_i)}, m_{2}^{(p_i)}, m_{3}^{(p_i)})$. 
Recall that each $m_{j}^{(p_i)}$ ($j = 1, 2, 3$) defines a 4-tuple of pairwise orthogonal 
projective lines in $A$. Redenote it (this 4-element set) by $T_{i, j}$. 
A (straightforward) Maple computation shows that 
it is possible to define $\alpha : \lbrace 1, 2, \dots, 9 \rbrace \to 
\lbrace 1, 2, 3 \rbrace$ in such a way, that the sets 
$\lbrace T_{i, \alpha (i)} \rbrace_{i = 1}^{9}$ are pairwise disjoint. 
In fact, there will be just 6 such functions $\alpha$; denote them as 
$\alpha_{1}, \alpha_{2}, \dots \alpha_{6}$. 
For each $k = 1, 2, \dots 6$, we have nine 4-tuples   
$\lbrace T_{i, \alpha_{k} (i)} \rbrace_{i = 1}^{9}$, and each $i$-th tuple can be extended in three 
different ways up to a 6-tuple (of pairwise orthogonal elements) by way of adjoining 
$\lbrace u_{p_i}, v_{p_i} \rbrace$, $\lbrace u_{p_i}', v_{p_i}' \rbrace$, or 
$\lbrace u_{p_i}'', v_{p_i}'' \rbrace$, respectively.

The third step is to try to implement a good bicolouring. 
For each $i = 1, 2, \dots, 9$ and $k = 1, 2, \dots, 6$ we have three elements of 
$\mathcal{P}_{\perp}^{(6)} (\wt{A})$ looking as follows: 
$T_{i, \alpha_k (i)} \cup \lbrace u_{p_i}, v_{p_i} \rbrace$, 
$T_{i, \alpha_k (i)} \cup \lbrace u_{p_i}', v_{p_i}' \rbrace$, and 
$T_{i, \alpha_k (i)} \cup \lbrace u_{p_i}'', v_{p_i}'' \rbrace$.  
Select any $i$ and $k$. 
If one assigns 1 (the red colour) to an element of $T_{i, \alpha_{k} (i)}$, this implies 
that all the corresponding $u$ and $v$ lines, as well as the rest of the lines in $T_{i, \alpha_{k} (i)}$, 
acquire the assignment 0 (the blue colour). 
On the other hand, if 1 (the red colour) is assigned to $u$ or $v$ element, say to $u_{p_i}$, then 
$v_{p_i}$ becomes blue (is assigned with 0), as well as the four elements of the 4-tuple. 
The latter implies that one of the elements in $\lbrace u_{p_i}', v_{p_i}' \rbrace$ and one of the elements in 
$\lbrace u_{p_i}'', v_{p_i}'' \rbrace$ should be red (i.e. have the label 1). 
In total, one obtains 12 (i.e. $4 + 2^3$) possible choices of colours for the 10 elements of 
\begin{equation*}
c_{i, k} := 
T_{i, \alpha_{k} (i)} \cup \lbrace u_{p_i}, v_{p_i}, u_{p_i}', v_{p_i}', u_{p_i}'', v_{p_i}'' \rbrace. 
\end{equation*}
A bicolouring $\varkappa : \wt{A} \to \lbrace 0, 1 \rbrace$ restricted to $c_{i, k}$ is a map 
$\varkappa^{(i, k)} : c_{i, k} \to \lbrace 0, 1 \rbrace$. 
We have $12$ candidates for $\varkappa^{(i, k)}$ in case $\varkappa$ is good; 
denote them $\varkappa_{1}^{(i, k)}, 
\varkappa_{2}^{(i, k)}, \dots \varkappa_{12}^{(i, k)}$. 
The corresponding sets of red lines 
$R_{l}^{(i, k)} := \lbrace x \in c_{i, k} \, | \, \varkappa_{l}^{(i, k)} (x) = 1 \rbrace$, 
$l = 1, 2, \dots, 12$, 
are either singletons, or 3-element sets. 
Recall that two red lines cannot be orthogonal. 
For each $k = 1, 2, \dots, 6$, 
put $D_{k} (i_1, l_1; i_2, l_2) := 1$, if $\forall x \in R_{l_1}^{(i_1, k)} 
\forall y \in R_{l_2}^{(i_2, k)}: y \not \perp x$; 
put $D_{k} (i_1, l_1; i_2, l_2) := 0$, -- otherwise 
($i_1, i_2 = 1, 2, \dots 9$; $l_1, l_2 = 1, 2, \dots 12$). 

A Maple computation shows that 
for each pair $(i_1, i_2)$, $i_1 < i_2$, there are 42 pairs $(l_1, l_2)$ such that 
$D_{k} (i_1, l_1; i_2, l_2) = 1$. 
Consider the set $L_k$ consisting of all tuples $(l_1, l_2, \dots, l_9)$ (where 
each $l_m$ ($m = 1, 2, \dots, 9$) is in the range $1, 2, \dots, 12$), such that 
$\forall m, n = 1, 2, \dots 9: D_{k} (m, l_m; n, l_n) = 1$. 
It turns out (Maple computation) that this set has just 5 elements, i.e. 
for each $k = 1, 2, \dots, 6$, 
there are just 5 ways to colour the elements $C_k := \cup_{i = 1}^{9} c_{i, k}$.    
Each of it's elements $l_{*} \equiv (l_1, l_2, \dots, l_9) \in L_{k}$ defines a collection of projective lines -- 
a subset $R_k (l_{*})$ of $C_k = \cup_{i = 1}^{9} c_{i, k}$ consisting of elements assigned with 1 (red).
The rest are assigned with 0 (blue). 

Let us fix at this point our achievements. 
We have constructed six subsets $C_k$ of $\wt{A}$ 
($k$ varies over $1, 2, \dots, 6$). If $\varkappa : \wt{A} \to \lbrace 0, 1 \rbrace$ is \emph{good}, 
then we can tell something about the restriction of $\varkappa$ to $C_k$: 
we have a limited number of options for $\varkappa |_{C_k}$ indexed by 
$l_{*} \in L_{k}$, $|L_k| = 5$. 

Recall that the set of red rays in $C_k$ corresponding to $l_{*} \in L_k$ is denoted by $R_k (l_{*})$, 
$k = 1, 2, \dots, 6$. 
For each $k$, consider the 9-element set 
$\wt{\alpha}_{k} := \lbrace \alpha_{k} (i) \rbrace_{i = 1}^{9}$. 
It turns out (Maple computation) that one can find triples $(q_1, q_2, q_3)$ 
(let $1 \leqslant q_1 < q_2 < q_3 \leqslant 9$) 
such that the corresponding 
$\wt{\alpha}_{q_1}$, $\wt{\alpha}_{q_2}$, $\wt{\alpha}_{q_3}$ are pairwise disjoint, i.e. 
their union has cardinality 27. 
Furthermore, there will be just 2 such triples $(q_1, q_2, q_3)$. 
Take any $l_{*} \in L_k$. 
One verifies that for each of the two possibilities of $(q_1, q_2, q_3)$ the set
$R_{q_1} (l_{*}) \cup R_{q_2} (l_{*}) \cup R_{q_3} (l_{*})$,  will be the same; 
denote it by $R_k (l_{*})$. 
Furthermore, the set of sets $\lbrace R_k (l_{*}) \rbrace_{l_{*} \in L_k}$ 
will be the same for each $k$; hence one may denote its five elements by 
$\mathcal{R}_{1}, \mathcal{R}_{2}, \dots, \mathcal{R}_{5}$. 
As a remark, two of them consist of 27 elements, and the other three have cardinalities 21. 
  
At this point we can say the following: if $\varkappa : \wt{A} \to \lbrace 0, 1 \rbrace$ is a good 
bicolouring, then the set of its red rays contains one of $\mathcal{R}_{m}$, $m = 1, 2, \dots, 5$. 

Now, for each $m = 1, 2, \dots, 5$, consider 
$\mathcal{B}_{m}$ consisting of all such lines $x \in \wt{A}$ (recall that we have them 198), for which 
there exists $y \in \mathcal{R}_{m}$ such that $y \perp x$. 
If $\mathcal{R}_m$ is contained in the set of red rays of $\varkappa$, then all elements of $\mathcal{B}_m$ 
must be blue (by the definition of good bicolouring). 
Now look at $\mathcal{P}_{\perp}^{(6)} (\wt{A})$. 
If $\mathcal{B}_m$, $m = 1, 2, \dots, 5$,  contains at least one of these 6-tuples, the 
corresponding variant with $m$ should be ruled out. 
It turns out (Maple computation) that just one of the five variants survives after the 
verification of this condition. 
Denote the set $\mathcal{R}_m$ corresponding to this unique $m$ by $\wt{\mathcal{R}}$. 
From a Maple computation we obtain, that $\wt{\mathcal{R}}$
consist of 27 lines of the shape  
$[0, 0, 0; \, 2, \bar 1, \bar 1; \, \bar 2, 1, 1]$, i.e. those which are represented by a vector with 
precisely three zeros. 
In the notation of the theorem, this is just the orbit $O (\lambda_4)$. 
It remains to check that if we colour all rays from $\wt{\mathcal{R}}$ to red, and all 
other rays in $\wt{A}$ to blue, then the conditions of the definition of a good bicolouring 
are satisfied. This yields the \emph{unique} good bicolouring of $\wt{A}$. 

Now let us consider the 
set $\wh{A} \supset \wt{A}$ from the theorem. 
As a side effect of the computations, 
the numbers $\wh{n}_k$ of elements in 
$\mathcal{P}_{\perp}^{(k)} (\wh{A})$, $k = 2, 3, \dots, 6$ are as follows: 
$\wh{n}_2 = 18423$, 
$\wh{n}_3 = 104978$, 
$\wh{n}_4 = 136620$, 
$\wh{n}_5 = 66744$, 
$\wh{n}_6 = 11124$. 
For each $l = [x_1, x_2, x_3; y_1, y_2, y_3; z_1, z_2, z_3] \in \wt{A}$, 
construct a ray $l_1 := [-x_1, -x_2, -x_3; y_1, y_2, y_3; z_1, z_2, z_3]$. 
Denote the union of all $l_1$ by $\wt{A}_1$. 
Similarly define the sets $\wt{A}_2$ and $\wt{A}_3$ as the unions of all $l_2$ of the form 
$l_2 := [x_1, x_2, x_3; -y_1, -y_2, -y_3; z_1, z_2, z_3]$, and all 
$l_3 := [x_1, x_2, x_3; y_1, y_2, y_3; -z_1, -z_2, -z_3]$, respectively. 

We have $\wh{A} \supset \wt{A}, \wt{A}_1, \wt{A}_2, \wt{A}_3$. 
Suppose that $\wh{A}$ admits a good bicolouring $\wh{\varkappa}$. 
The red subset $\wt{\mathcal{R}}$ of $\wt{A}$ 
is known (the 27 elements of the ortbit $O (\lambda_4)$). 
Consider three reflections $P_1 : [x; y, z] \mapsto [-x; y; z]$, 
$P_2 : [x; y, z] \mapsto [x; -y; z]$, 
$P_3 : [x; y, z] \mapsto [x; y; -z]$. 
The red subsets of $\wt{A}_i$ (to be denoted $\wt{\mathcal{R}}_i$), $i = 1, 2, 3$, are obtained by applying 
these reflections to $\wt{\mathcal{R}}$. 
Put $\wh{\mathcal{R}} := \wt{\mathcal{R}} \cup \wt{\mathcal{R}}_1 \cup \wt{\mathcal{R}}_2 \cup \wt{\mathcal{R}}_3$. 
The cardinality of $\wh{\mathcal{R}}$ will be 54. 
The set of red rays of $\wh{\varkappa}$ should contain $\wh{\mathcal{R}}$. 
The non-colourability of $\wh{A}$ is derived now from the following fact (checked in Maple): 
there exists a 6-tuple 
from $\mathcal{P}_{\perp}^{(6)} (\wh{A})$ such that 
the cardinality of it's intersection with $\wh{\mathcal{R}}$ is not equal to 1, i.e. 
either we obtain a completely blue 6-tuple of pairwise orthogonal lines or 
encounter a situation when several red lines are mutually orthogonal. 
This contradicts the assumption that $\wh{\varkappa}$ is good. Therefore $\wh{A}$ is non-colourable.

\section{Discussion.}

It is interesting to mention that 
the notion of a saturated projective configuration is intimately related with the notion of an orthoalgebra. 
An \emph{orthoalgebra} is a set $S$ equipped with a relation $\perp \subset S \times S$, a map 
$\cdot \oplus \cdot: \perp \to S$, $(x, y) \mapsto x \oplus y$, and two distinct elements 
$\mathbf{0}, \mathbf{1} \in S$; these data satisfy  
(1) if $x \oplus y$ is defined, then $x \oplus y = y \oplus x$; 
(2) if $(x \oplus y) \oplus z$ is defined, then $(x \oplus y) \oplus z = x \oplus (y \oplus z)$; 
(3) $x \oplus \mathbf{0}$ is always defined and $x \oplus \mathbf{0} = x$;  
(4) $\forall x \, \exists ! x^{*}: x \oplus x^{*} = \mathbf{1}$; 
(5) if $x \oplus x$ is defined, then $x = \mathbf{0}$.  
A prototypical example of an orthoalgebra is the Hilbert space orthoalgebra $\mathbb{L} (\mathcal{H})$: 
the set $S$ is the set of all subspaces of the Hilbert space $\mathcal{H}$, and $\oplus$ is 
the orthogonal sum.  

If $A$ is a finite saturated projective configuration in $\mathcal{H}$, then it generates a finite 
suborthoalgebra of $\mathbb{L} (\mathcal{H})$. 
The examples of such configurations are given above, but there exist others, for instance, 
\cite{Peres}, \cite{ZimbaPenrose}.  
Note, that the corresponding partial Boolean algebra 
(see \cite{Smith}), need not be finite. 
Orthoalgebras attract attention as the structures capturing the logic of quantum theory 
\cite{Wilce}, \cite{Isham}. 
The relation between ``quantum logic'' and Kochen-Specker-type constructions (i.e. non-bicolourable 
finite configurations) is discussed in \cite{Meyer}, \cite{Kent} and in  
\cite{SvozilTkadlec}, \cite{IshamButterfield}. 
If the finite saturated configuration $A$ is non-bicolourable (i.e. is of Kochen-Specker type), then 
this fact is translated into the absence of a morphism from the corresponding orthoalgebra to a 
two-element orthoalgebra (absence of bivaluations). 
A series of examples of such orthoalgebras has been constructed in \cite{RuugeFVO2}; 
in particular, the orthoalgebra corresponding to the configuration described in \cite{RuugeFVO1} is 
isomorphic to the $E_8$ orthoalgebra.

\vspace{0.3 true cm}
This work has been financially supported by FWO (Belgium).


\begin{thebibliography}{99}

\bibitem{KochenSpecker}
Kochen, S.; Specker, E. P. ``The problem of hidden variables in quantum mechanics.''  
\emph{J. Math. Mech.}  \textbf{17} (1967), 59--87.

\bibitem{Peres} 
Peres, A. ``Two simple proofs of the Kochen-Specker theorem.''  
\emph{J. Phys. A}  \textbf{24}  (1991), no. 4, L175--L178.

\bibitem{Cabello}
Cabello, A.; Estebaranz, M.J.; García-Alcaine, G. 
``Bell-Kochen-Specker theorem: a proof with $18$ vectors.''  
\emph{Phys. Lett. A} \textbf{212}  (1996),  no. 4, 183--187. 

\bibitem{Mermin}
Mermin, N.D. ``Hidden variables and the two theorems of John Bell.''  
\emph{Rev. Modern Phys.}  \textbf{65}  (1993),  no. 3, part 1, 803--815.

\bibitem{KernaghanPeres}
Kernaghan, M.; Peres, A. ``Kochen-Specker theorem for eight-dimensional space.''  
\emph{Phys. Lett. A}  \textbf{198}  (1995),  no. 1, 1--5.


\bibitem{RuugeFVO1}
Ruuge, A.E.; Van Oystaeyen, F. ``Saturated Kochen-Specker-type configuration of 120 projective 
lines in eight-dimensional space and its group of symmetry.''  
\emph{J. Math. Phys.}  \textbf{46} (2005),  no. 5, 052109, 28 pp.


\bibitem{Aravind}
Aravind, P.K.; Lee-Elkin, F. 
``Two noncolourable configurations in four dimensions illustrating the Kochen-Specker theorem.''  
\emph{J. Phys. A}  \textbf{31}  (1998),  no. 49, 9829--9834.

\bibitem{Smith}
Smith, D. ``Algebraic partial Boolean algebras.''  
\emph{J. Phys. A}  \textbf{36}  (2003),  no. 13, 3899--3910.

\bibitem{ZimbaPenrose}
Zimba, J.; Penrose, R. ``On Bell nonlocality without probabilities: more curious geometry.'' 
\emph{Stud. Hist. Philos. Sci.} \textbf{24} (1993), 697--720.  

\bibitem{Wilce}
 Wilce, A. ``Compact orthoalgebras.''  
\emph{Proc. Amer. Math. Soc.}  \textbf{133}  (2005),  no. 10, 2911--2920. 

\bibitem{Isham}
Isham, C. J. ``Topos theory and consistent histories: the internal logic of the set of all consistent sets.''  
\emph{Internat. J. Theoret. Phys.}  \textbf{36}  (1997),  no. 4, 785--814.

\bibitem{Meyer} 
Meyer, D.A. ``Finite precision measurement nullifies the Kochen-Specker theorem.''  
\emph{Phys. Rev. Lett.}  \textbf{83}  (1999),  no. 19, 3751--3754.

\bibitem{Kent}
Kent, A. ``Noncontextual hidden variables and physical measurements.''  
\emph{Phys. Rev. Lett.}  \textbf{83}  (1999),  no. 19, 3755--3757.


\bibitem{SvozilTkadlec}
Svozil, K.; Tkadlec, J. 
``Greechie diagrams, nonexistence of measures in quantum logics, and Kochen-Specker-type constructions.''  
\emph{J. Math. Phys.}  \textbf{37}  (1996),  no. 11, 5380--5401. 

\bibitem{IshamButterfield}
Isham, C. J.; Butterfield, J. ``Topos perspective on the Kochen-Specker theorem. I. 
Quantum states as generalized valuations.''  
\emph{Internat. J. Theoret. Phys.}  \textbf{37}  (1998),  no. 11, 2669--2733. 



\bibitem{RuugeFVO2}
Ruuge, A.E.; Van Oystaeyen, F. ``New families of finite coherent orthoalgebras without bivaluations.''  
\emph{J. Math. Phys.}  \textbf{47}  (2006),  no. 2, 022108, 32 pp.


\end{thebibliography}
\end{document}